\documentclass[a4paper]{eslab}
\usepackage{graphics}

\def\apj#1{{ApJ} {#1}}
\def\apjs#1{{ApJS} {#1}}
\def\aa#1{{A\&A} {#1}}

\def\mnras#1{{MNRAS} {#1}}
\def\araa#1{{ARA\&A} {#1}}
\def\apss#1{{Ap\&SS} {#1}}

\def\be{\begin{equation}}
\def\ee{\end{equation}}

\def\bea{\begin{eqnarray}}
\def\eea{\end{eqnarray}}

\def\eg{~e.g.~}
\def\ie{~i.e.~}
\def\etal{~et al.~}

\title{Chemo - Photometric evolution of star forming disk galaxy}

\author{Peter Berczik}

\institute{Main Astronomical Observatory of Ukrainian National Academy 
of Sciences, \\ UA-03680, Golosiiv, Kiev-127, Ukraine, e-mail: {\tt 
berczik@mao.kiev.ua}}

\begin{document}

\maketitle 


\begin{abstract}

The chemical and photometric evolution of star forming disk galaxies is 
investigated. Numerical simulations of the complex gasdynamical flows 
are based on our own coding of the Chemo - Dynamical Smoothed Particle 
Hydrodynamical (CD - SPH) approach, which incorporates the effects of 
star formation. As a first application, the model is used to describe 
the chemical and photometric evolution of a disk galaxy like the
Milky Way. 


\end{abstract}


\section{Introduction}

Galaxy formation is a highly complex subject 
requiring many different approaches of investigation. Recent advances in computer 
technology and numerical methods have allowed detailed modeling of 
baryonic matter dynamics in a universe dominated by collisionless dark 
matter and, therefore, the detailed gravitational and hydrodynamical 
description of galaxy formation and evolution. The most sophisticated 
models include radiative processes, star formation and supernova 
feedback (\eg \cite{K92,SM94,FB95}).

The results of numerical simulations are fundamentally affected by the 
star formation algorithm incorporated into modeling techniques. 
Yet star formation and related processes are still not well understood 
on either small or large spatial scales. Therefore the star formation 
algorithm by which gas is converted into stars can only be based on 
simple theoretical assumptions or on empirical observations of nearby 
galaxies. 

Among the numerous methods developed for modeling complex three 
dimensional hydrodynamical phenomena, Smoothed Particle Hydrodynamics 
(SPH) is one of the most popular (\cite{M92}). Its Lagrangian nature 
allows easy combination with fast N - body algorithms, making  
 possible the simultaneous description of complex gas-stellar
dynamical systems (\cite{FB95}). As an example of such a combination, an TREE - SPH 
code (\cite{HK89,NW93}) was successfully applied to the detailed modeling 
of disk galaxy mergers (\cite{MH96}) and of galaxy formation and evolution 
(\cite{K92}). A second good example is an GRAPE - SPH code (\cite{SM94,SM95}) 
which was successfully used to model the evolution of disk galaxy 
structure and kinematics.

\begin{figure}[!ht]
\resizebox{3.3in}{!}{\includegraphics{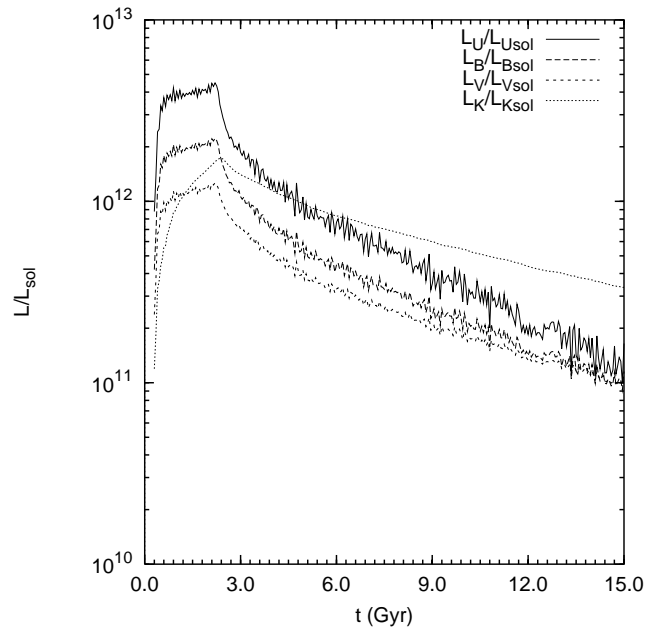}}
\caption{Luminosity evolution of the model galaxy. 
         \label{lum-t}}
\end{figure}

\begin{figure}[!ht]
\resizebox{3.3in}{!}{\includegraphics{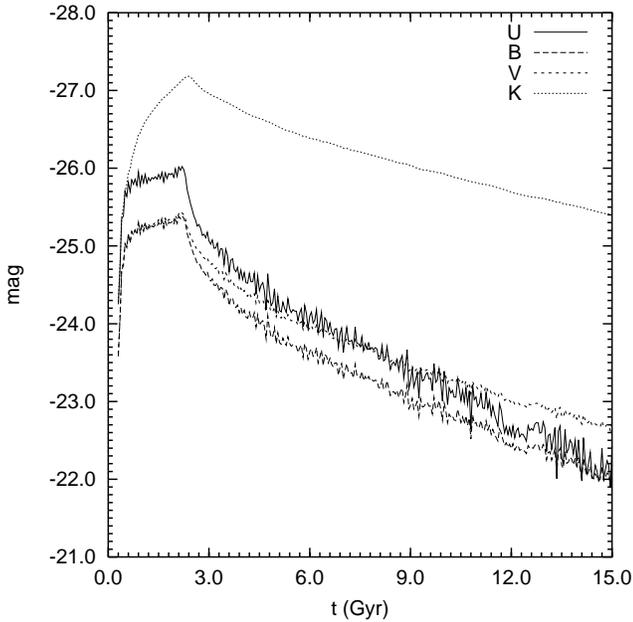}}
\caption{Photometric evolution of the model galaxy. 
         \label{ubvk-t}}
\end{figure}

\begin{figure}[!ht]
\resizebox{3.3in}{!}{\includegraphics{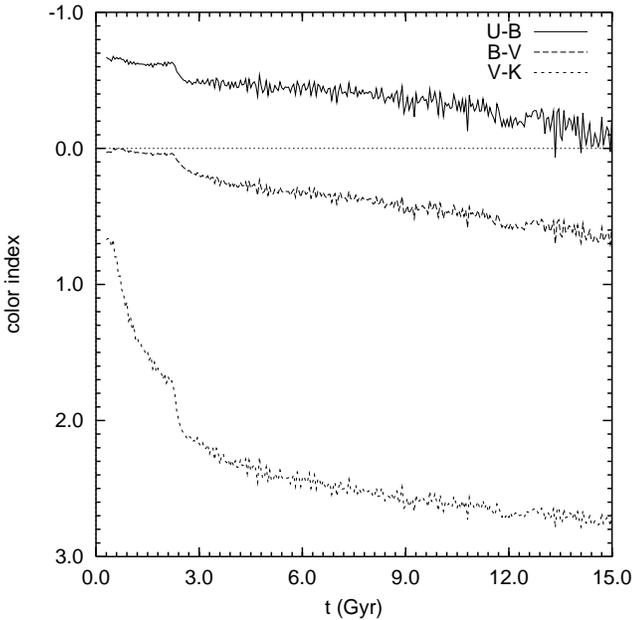}}
\caption{Color index evolution of the model galaxy. 
         \label{ci-t}}
\end{figure}


\section{Model}

The hydrodynamical simulations are based on our own coding of the 
Chemo - Dynamical Smoothed Particle Hydrodynamics (CD - SPH) approach, 
including feedback through star formation (SF). The dynamics of the "star" 
component is treated in the frame of a standard N - body approach. Thus, 
the galaxy consists of "gas" and "star" particles. For a detailed 
description of the CD - SPH code (the star formation algorithm, the 
SNII, SNIa and PN production, the chemical enrichment and the initial 
conditions) the reader is referred to \cite*{BerK96,Ber99}. Here we
briefly decsribe the basic features of our algorithm.

We modify the standard SPH SF algorithm (\cite{K92,SM94,SM95}), taking 
into account the presence of chaotic motion in the gaseous environment and 
the time lag between the initial development of suitable conditions for SF, and SF itself.
   
Inside a "gas" particle, the SF can start if the absolute value of the "gas" particle 
gravitational energy exceeds the sum of its thermal energy and energy of chaotic 
motions:
   
   \be
   \mid E_i^{gr} \mid > E_i^{th} + E_i^{ch}.
   \ee 

The chosen "gas" particle produces stars only if the above condition holds over 
the time interval exceeding its free - fall time:
   
   \be 
   t_{ff} = \sqrt { \frac{3 \cdot \pi}{32 \cdot G \cdot \rho} }.
   \ee 
   
We also check that the "gas" particles remain cool, \ie $ t_{cool} < t_{ff} $. 
We rewrite these conditions following \cite*{NW93}:
   
   \be 
   \rho_i > \rho_{crit}.
   \ee 
   
We set the value of $\rho_{crit} = 0.03$ cm$^{-3}$.
   
When the collapsing particle $ i $ is defined, we create the new 
"star" particle with mass $ m^{star} $ and update the "gas" particle
$ m_i $ using these simple equations: 
   
   \bea
   \left\{
   \begin{array}{lll}
   m^{star} = \epsilon  \cdot m_i,  \\
                                   \\
   m_i = (1 - \epsilon) \cdot m_i.  \\
   \end{array}
   \right.
   \eea

In the Galaxy, on the scale of giant molecular clouds, the typical 
values for SF efficiency are in the range $ \epsilon \approx 0.01 
\div 0.4 $ (\cite{DIL82,WL83}).
   
We did not fixe this value but rather also derived $ \epsilon $ from the 
"energetics" condition:
   
   \be 
   \epsilon = 1 - \frac{E_i^{th} + E_i^{ch}}{\mid E_i^{gr} \mid}.
   \ee
   
At the moment of birth, the positions and velocities of new 
"star" particles are set equal to those of parent "gas" particles. 
Thereafter these "star" particles interact with other "gas" and 
"star" or "dark matter" particles only by gravity. 

\begin{figure}[!ht]
\resizebox{3.3in}{!}{\includegraphics{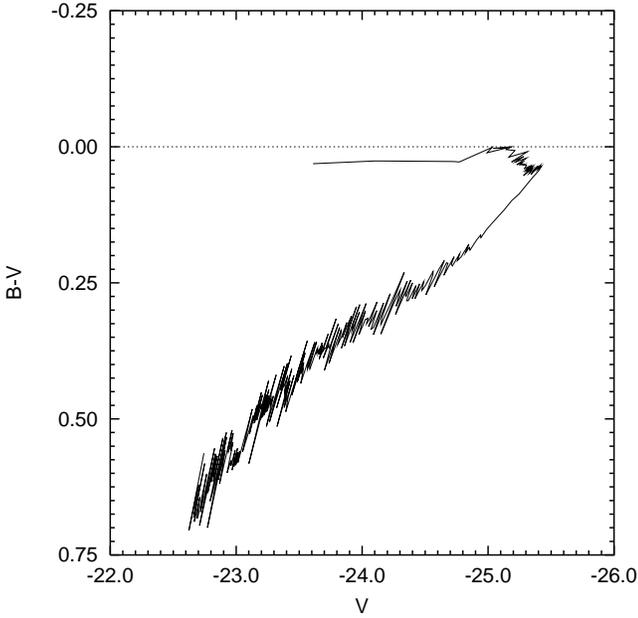}}
\caption{Evolution of the model galaxy in the (B-V) vs. V plane. 
         \label{bv-v}}
\end{figure}

\begin{figure}[!ht]
\resizebox{3.3in}{!}{\includegraphics{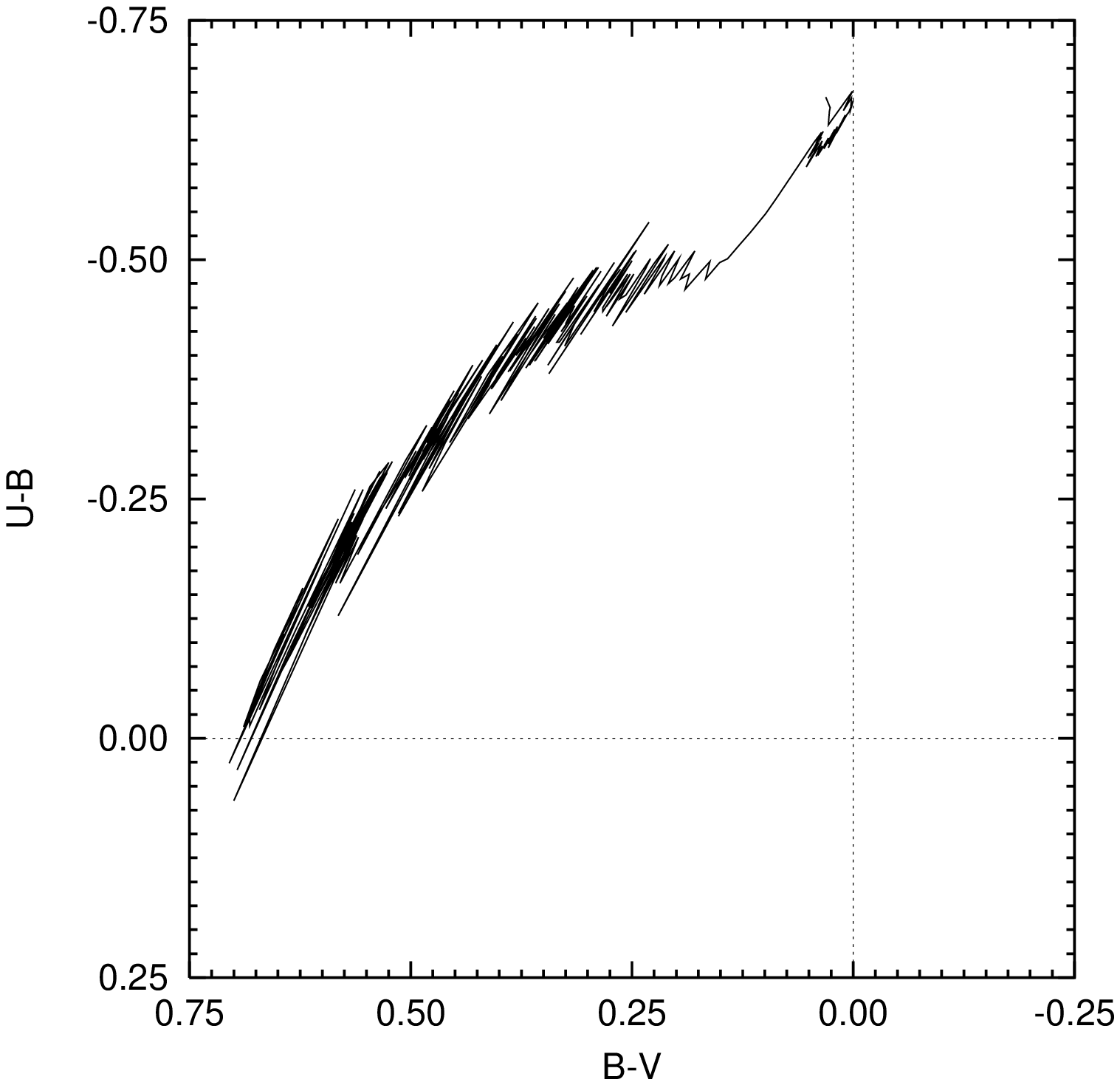}}
\caption{Evolution of the model galaxy in the (U-B) vs. (B-V) plane. 
         \label{ub-bv}}
\end{figure}

\begin{figure}[!ht]
\resizebox{3.3in}{!}{\includegraphics{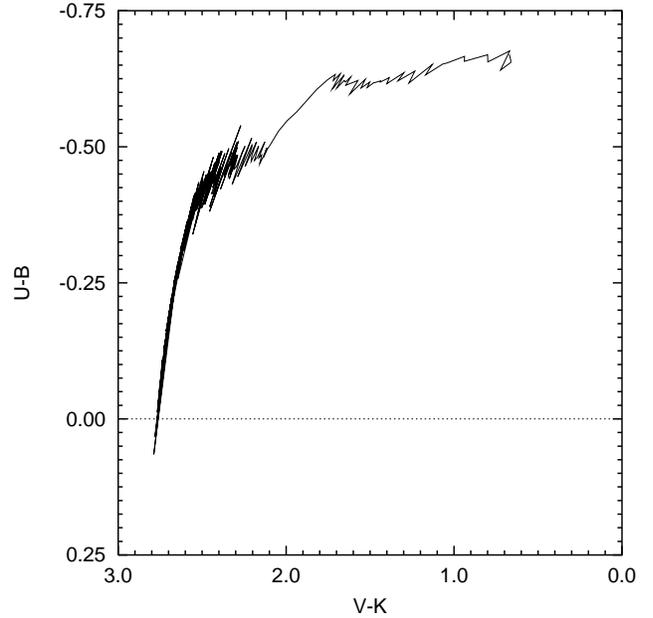}}
\caption{Evolution of the model galaxy in the (U-B) vs. (V-K) plane.
         \label{ub-vk}}
\end{figure}

For the thermal budget of the ISM, SNIIs play the  main role. Following 
to \cite*{K92,FB95}, we assume that the energy from the explosion is converted totally 
to thermal energy. The total energy released by SNII explosions 
($ 10^{44} \; J $ per SNII) within "star" particles is calculated at each 
time step and distributed uniformly between the surrounding "gas" particles 
(\cite{RVN96}).
   
In our SF scheme, every new "star" particle represents a separate, 
gravitationally bound, star formation macro region (like a 
globular clusters). The "star" particle has its own time of birth 
$ t_{begSF} $ which is set equal to the moment the particle is
formed. After the formation, these particles return the chemically enriched 
gas into surrounding "gas" particles due to SNII, SNIa and PN events.

We concentrate our treatment only on the production of $^{16}$O and 
$^{56}$Fe, yet attempt to describe the full galactic time evolution of 
these elements, from the beginning up to present time (\ie $ 
t_{evol} \approx 15.0 $ Gyr).

The code also includes the photometric evolution of each "star" 
particle, based on the idea of the Single Stellar Population (SSP) 
(\cite{BCF94,TCBF96}). 

At each time - step, absolute magnitudes: M$_U$, M$_B$, M$_V$, M$_R$, M$_I$, 
M$_K$, M$_M$ and M$_{bol}$ are defined separately for each "star" 
particle. The SSP integrated colours (UBVRIKM) are taken from 
\cite*{TCBF96}. The spectro - photometric evolution of the 
overall ensemble of "star" particles forms the Spectral Energy 
Distribution (SED) of the galaxy.

We do not model the energy distribution in spectral lines nor the 
scattered light by dust. However according to \cite*{TCBF96} our 
approximation is reasonable, especially in the UBV spectral brand. 


\section{Results}

The model presented descibes well the time evolution of the basic 
chemical and photometric parameters of a disk galaxy similar to the
Milky Way. The metallicity, luminosity and colors obtained are
typical of such disk galaxies.

\begin{figure}[!ht]
\resizebox{3.3in}{!}{\includegraphics{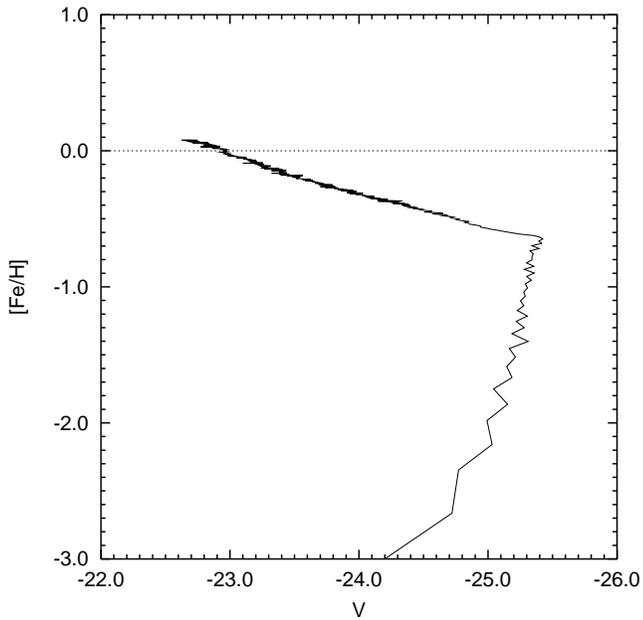}}
\caption{Evolution of the model galaxy in the [Fe/H] vs. V plane.
         \label{fe-v}}
\end{figure}

\begin{figure}[!ht]
\resizebox{3.3in}{!}{\includegraphics{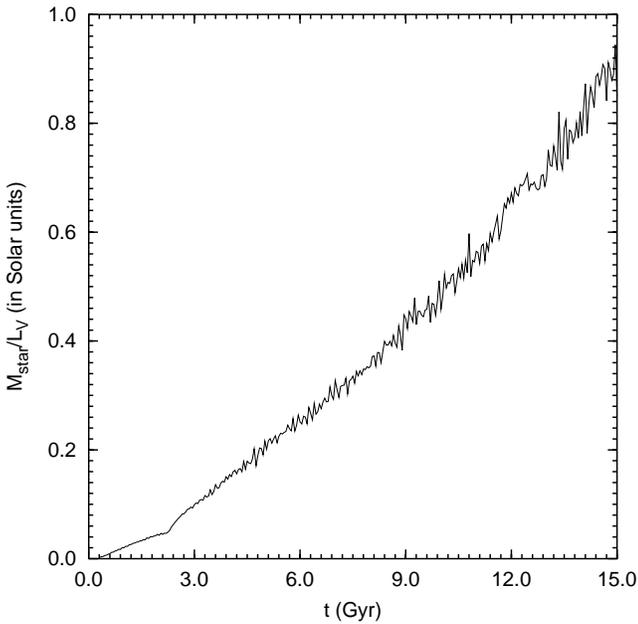}}
\caption{The "real" M$_{star}$/L$_{V}$ evolution of the model galaxy.
         \label{ml-t}}
\end{figure}

\begin{itemize}

   \item Figure~\ref{lum-t}. Luminosity evolution of the model galaxy. 
   
   \item Figure~\ref{ubvk-t}. Photometric evolution of the model galaxy.
   
   \item Figure~\ref{ci-t}. Color index evolution of the model galaxy.
   
   \item Figure~\ref{bv-v}. Evolution of the model galaxy in the (B-V) 
         vs. V plane.
         
   \item Figure~\ref{ub-bv}. Evolution of the model galaxy in the (U-B) 
         vs. (B-V) plane.
   
   \item Figure~\ref{ub-vk}. Evolution of the model galaxy in the (U-B) 
         vs. (V-K) plane.
         
   \item Figure~\ref{fe-v}. Evolution of the model galaxy in the [Fe/H] 
         vs. V plane.
         
   \item Figure~\ref{ml-t}. The "real" M$_{star}$/L$_{V}$ evolution of 
         the model galaxy.
   
\end{itemize}

\begin{acknowledgements}

The author is grateful to Yu.I. Izotov and L.S. Pilyugin for fruitful 
discussions during the process of preparing this work. 

The work was supported by the German Science Foundation (DFG)
under grants No. 436 UKR 18/2/99, 436 UKR 17/11/99 and partially 
supported by NATO grant NIG 974675. 

Special thanks for hospitality to the Astronomisches Rechen-Institute 
(ARI) Heidelberg, where part of this work has been done. 

It is a pleasure to thank Christian Boily for comments on an earlier 
version of this work.

\end{acknowledgements}



\end{document}